\documentclass[twocolumn,superscriptaddress,aps,prb]{revtex4-1}
\usepackage{graphicx}
\usepackage{graphicx,amsfonts}
\usepackage{color}
\usepackage{float}
\usepackage{bm}
\begin{document}

\title{Probing the superconducting gap symmetry of $\alpha$-PdBi$_{2}$: A penetration depth study}
\author{S. Mitra}
\affiliation{Division of Physics and Applied Physics, School of Physical and Mathematical Sciences, Nanyang Technological University, 21 Nanyang Link, Singapore 637371}
\author{K. Okawa}
\affiliation{Laboratory for Materials and Structures
Laboratory, Tokyo Institute of Technology, Kanagawa 226-8503, Japan}
\author{S. Kunniniyil Sudheesh}
\affiliation{Division of Physics and Applied Physics, School of Physical and Mathematical Sciences, Nanyang Technological University, 21 Nanyang Link, Singapore 637371}
\author{T. Sasagawa}
\affiliation{Laboratory for Materials and Structures
Laboratory, Tokyo Institute of Technology, Kanagawa 226-8503, Japan}
\author{Jian-Xin Zhu}
\affiliation{Theoretical Division and Center for Integrated Nanotechnologies, Los Alamos National Laboratory, Los Alamos, New Mexico 87545, USA}
\author{Elbert E. M. Chia}
\email{elbertchia@ntu.edu.sg}
\affiliation{Division of Physics and Applied Physics, School of Physical and Mathematical Sciences, Nanyang Technological University, 21 Nanyang Link, Singapore 637371}
\date{\today}

\begin{abstract}
We report measurements of the in-plane London penetration depth $\lambda$ in single crystals of the $\alpha$-PdBi$_{2}$ superconductor --- the $\alpha$-phase counterpart of the putative topological superconductor $\beta$-PdBi$_{2}$, down to 0.35~K using a high-resolution tunnel-diode-based technique. Both $\lambda$ and superfluid density $\rho_{s}$ exhibit an exponential behavior for $T\leq$ 0.35$T_{c}$, with $\Delta(0)/k_{B}T_{c}\sim$2.0, $\Delta C/\gamma T_{c}$$\sim$2.0 and $\lambda(0)$$\sim$140~nm, showing that $\alpha$-PdBi$_{2}$ is a moderately-coupling, fully-gapped superconductor. The values of $\Delta(0)$ and $\Delta C/\gamma T_{c}$ are consistent with each other via strong-coupling corrections. 

\end{abstract}

\maketitle

The recently discovered superconductor $\beta$-PdBi$_{2}$ ($T_{c}$$\sim$5.3~K) \cite{Sasagawa15} has been 
proposed as a possible candidate to exhibit topological superconductivity.  A topological superconductor (TSC) has zero-energy localized modes in its quasiparticle excitation spectrum called Andreev bound states at the surface, or Majorana fermions at the vortex core center, which are topologically protected. In the context of superconductivity this means that, a TSC is characterized by a fully-gapped bulk while these Majorana dispersing states can exhibit gapless excitation. Spin- and angle-resolved photoemission spectroscopy (ARPES) revealed 
the existence of several topologically protected surface states crossing the Fermi level in $\beta$-PdBi$_{2}$,\cite{Sasagawa15} though the experimental detection of Majorana fermions is still elusive.\cite{Liqiang16} Preliminary low-temperature (down to 2~K) specific heat measurements \cite{Imai12} hinted towards the possibility of a multi-gap superconducting phase in $\beta$-PdBi$_{2}$, while scanning tunneling microscopy (STM) \cite{Herrera15} suggested that it behaves like a single-gap multi-band superconductor. However, later experiments using muon-spin relaxation ($\mu$SR) \cite{Biswas16} and calorimetric studies \cite{Samuely16} have shown a single isotropic BCS-like gap in $\beta$-PdBi$_{2}$ with negligible contribution from the topologically protected surface states. 

Another extensively-researched superconducting compound amongst the Pd-Bi binary systems is $\alpha$-PdBi ($T_{c}$$\sim$3.7~K), that has a monoclinic crystal structure and belongs to the space group P$2_{1}$.\cite{Zhuravlev57,Bhatt79,Joshi11} Unlike $\beta$-PdBi$_{2}$  that is centrosymmetric, $\alpha$-PdBi lacks a center of inversion i.e. it exhibits noncentrosymmetric (NCS) superconductivity. In NCS superconductors, Rashba-type antisymmetric spin-orbit coupling is allowed,\cite{Rashba01} which is expected to lift the spin degeneracy and lead to a complex pairing wave function that might be characterized by a hybrid pairing of both spin-singlet and spin-triplet superconductivity.\cite{Joshi11,Jiao14} According to a recent review article by Smidman \textit{et al.}  
\cite{Smidman16}, a well-defined signature for distinguishing between the singlet and triplet component in the mixture should be the existence of topological states, which is exclusive to the pure spin-triplet pairing symmetry. However, it should be noted that experimental signature for spin-triplet superconductivity has also been observed in samples \cite{Spinttriplet98,Spinttriplet14} that do not posses any topologically non-trivial states. Interestingly, recent ARPES measurements \cite{Neupane16} on $\alpha$-PdBi revealed the presence of spin-polarized surface states at high binding energies but not at the Fermi level, thus negating the possibility of topological superconductivity at the surface. Scanning tunneling spectroscopy (STS) measurements hinted towards a moderately-coupled BCS-like single gap scenario in $\alpha$-PdBi, \cite{Sun15} similar to that reported for $\beta$-PdBi$_{2}$. 

\begin{figure}[h] 
	\centering \includegraphics[width=8cm,clip]{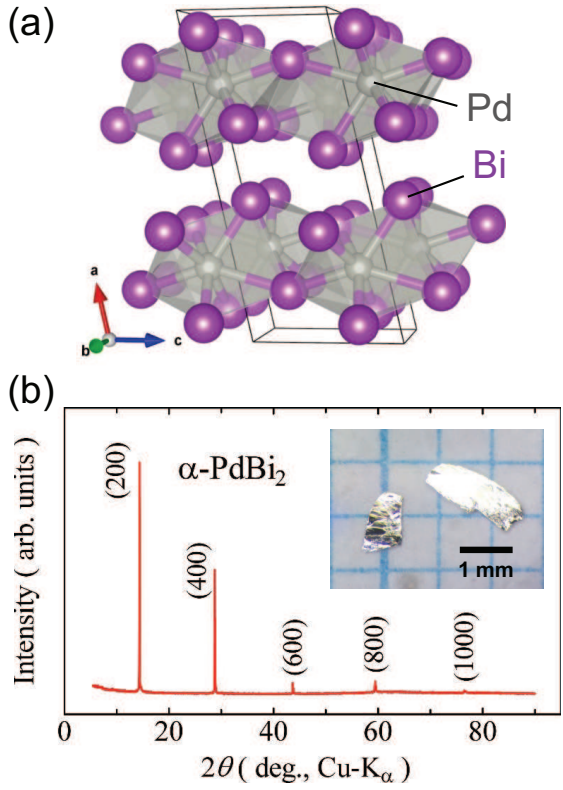}
	\caption{(a) Crystal structure of $\alpha$-PdBi$_{2}$. (b) x-ray diffraction pattern from the cleavage plane of the $\alpha$-PdBi$_{2}$ single crystal as shown in the inset. }
	\label{fig:XRD}
\end{figure}

We thus see that the presence of topological states has been consistently predicted in the Pd-Bi family of superconductors, even though experimental observation of topological superconductivity is yet to be confirmed. Since $\sim$2010, there has been a consistent effort to realize topological superconductors by carrier doping, e.g. Cu- and Nb-intercalated Bi$_{2}$Se$_{3}$,\cite{Hor10,Kriener11,Smylie16} and In-doped SnTe.\cite{Sasaki12} In contrast, the Pd-Bi family of binary compounds provide the opportunity for studying some of the first candidates for stoichiometric topological superconductors.\cite{Guan16} Along this line, the less-explored superconductor $\alpha$-PdBi$_{2}$ ($T_{c}$$\sim$1.7~K),\cite{Zhuravlev57} which is a structural isomer of $\beta$-PdBi$_{2}$, is
interesting to investigate. PdBi$_{2}$ has two distinct crystallographic phases --- the low-temperature $\alpha$-phase is obtained below 380$^\circ$C with
slow cooling, while the high-temperature $\beta$-phase can be
stabilized at low temperatures by rapid quenching between 380$^\circ$C to 490$^\circ$C.\cite{Matthia63,Okamoto94} The $\alpha$-PdBi$_{2}$ has a centrosymmetric monoclinic crystal structure of space group C2/m as
shown in Figure~\ref{fig:XRD}(a), while the $\beta$-PdBi$_{2}$ has a tetragonal structure belonging to the space group I4/mmm.\cite{Imai12,Xu92,Shein13}

Single crystals of $\alpha$-PdBi$_{2}$ were grown by a melt growth technique. Elemental Pd (3N5) and Bi (5N) at a molar ratio of 1:2 were sealed in an evacuated quartz tube, pre-reacted at high temperature until it completely melted and mixed. Then, it was again heated up to 900$^\circ$C, kept for 20 hours, cooled down slowly at a rate of 2--3$^\circ$C/h down to room temperature. The obtained single crystals by the optimized growth conditions had a good cleavage, producing flat surfaces as shown in the inset of Figure~\ref{fig:XRD}(b). The peaks of the x-ray diffraction from the cleavage plane can be assigned to the (\textit{h} 0 0) reflections (Figure~\ref{fig:XRD}(b)), indicating that the cleavage plane is the \textit{bc}-plane.   

\begin{figure}[h] 
	\centering \includegraphics[width=8cm,clip]{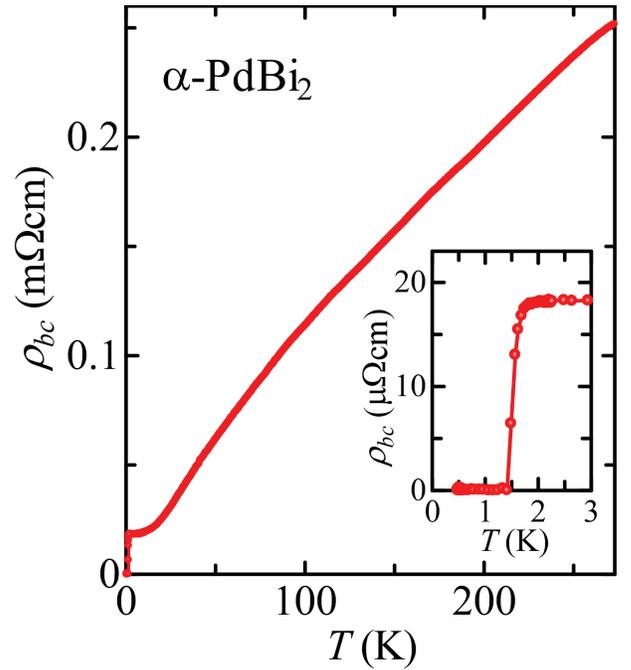}
	\caption{Resistivity versus temperature data of $\alpha$-PdBi$_{2}$, showing a $T_{c}$$\sim$1.7~K.}
	\label{fig:RT}
\end{figure} 

Resistivity in the \textit{bc}-plane of the $\alpha$-PdBi$_{2}$ crystal was measured by the four-probe method using the Keithley 2182A Nanovoltmeter and 6221 Current Source. A homemade adiabatic demagnetization refrigerator was used for temperature below 2~K. Temperature dependence of resistivity of $\alpha$-PdBi$_{2}$ in the wide temperature range and around the superconducting transition is shown in Fig.~\ref{fig:RT} and its inset. The residual resistivity below 10~K was adequately low (18~$\mu \Omega$-cm) and its ratio to the room temperature value (RRR: residual resistivity ratio) is 15, indicating the high quality of the crystal. The onset of the superconducting transition is $\sim$1.7~K.

 The tunnel-diode-oscillator (TDO) based penetration depth setup has been shown to be an excellent tool to probe the pairing symmetry of unconventional superconductors such as ruthenates, \cite{Bonalde2000} skutterudites, \cite{Chia03PRL,Chia04,Chia05} and pnictides, \cite{Matsuda12,Matsuda15,Prozorov12} due to its ability to discern very small changes (1 part in $10^9$) at low temperatures. At low temperatures, isotropic superconducting gaps give an exponential temperature dependence of the penetration depth, whereas nodes in the gap function, whether point nodes or line nodes, give a power-law temperature dependence. Coupled with the fact that penetration depth measurements are more surface-sensitive than bulk measurements, gapless excitations from the surface states of TSCs may be observable using the TDO technique, thus confirming the presence or absence of the topological nature of superconductivity in this material. In this paper, we present high precision measurements of the in-plane London penetration depth $\lambda$(\textit{T}) of $\alpha$-PdBi$_{2}$ down to 0.35~K using a TDO-based technique. The change in penetration depth $\Delta \lambda $(\textit{T}) shows an exponential behavior at low-temperatures, suggesting the presence of a single isotropic gap in this material. The best fit to the normalized in-plane superfluid density $\rho_{s}$(\textit{T}) is obtained for the zero-temperature superconducting gap $\Delta$(0)$\sim$2.0\textit{k}$_{B}$\textit{T}$_{c}$, and the specific heat jump $\Delta C/\gamma T_{c}$$\sim$2.0, where $\gamma$ is the electronic specific heat coefficient. This suggests that $\alpha$-PdBi$_{2}$ is a moderate-coupling, fully-gapped superconductor. Also, we do not see any power-law low-temperature dependence of $\Delta \lambda $(\textit{T}). This, however, is not definite evidence of the lack of gapless excitations on the surface of the sample, since value of the zero-temperature penetration depth $\lambda(0)$ is a few times the surface state thickness ($\sim$20-60~nm) in this material.\cite{Biswas16,Sun15}

The data presented here were taken on single crystal samples in the shape of platelets with dimensions $\sim$0.8 $\times$ 0.5 $\times$ 0.1 mm$^{3}$, the smallest dimension being oriented along the \textit{a}-axis. Measurements were performed using a tunnel diode oscillator \cite{Prozorov00,Prozorov01,Chia03} operating at a resonant frequency of 26~MHz. The system has been optimized to have a noise level of 2 parts in 10$^{9}$ with low drift ($\sim$0.02 Hz/minute). The cryostat dewar was surrounded by a bilayer Mumetal jacket to shield DC stray fields down to a few mOe.  Measurements were carried on using a Helium-3 cryostat (Cryogenics Industries of America), which is capable of cooling down our sample to 0.35~K. The sample is mounted using GE Varnish on a single crystal sapphire rod that is thermally connected (with silver epoxy) to a 99.999\% pure gold-plated Oxygen-Free-High-Conductive Copper cold finger, that is thermally anchored to the cryostat sample mount. A pre-calibrated Cernox-1030 temperature sensor from Lakeshore Cryogenics mounted at the base of the sapphire rod is capable of monitoring the temperature from the base temperature to 20~K.

Below the superconducting transition temperature $T_{c}$, the change in the London penetration depth $\Delta \lambda (T)$ causes a change in the magnetic susceptibility which in turn changes the inductance of the resonator coil, hence changing the resonant frequency $\Delta f (T)$.\cite{Prozorov00} It can be shown that $\Delta \lambda (T) = \lambda (T) - \lambda $(0.35~K) is directly related to $\Delta f(T)$ as 
$\Delta \lambda (T) = G \Delta f (T)$. Here $G$ is a proportionality factor that depends on the coil and sample geometries. We first obtain $G$ for a 99.9995\% pure Aluminum single crystal (of known dimensions) by adjusting \textit{G} until the normalized superfluid data fits the extreme non-local BCS expression. We then estimate \textit{G} for our sample of known dimensions. This technique works particularly well for samples with regular dimensions, large aspect ratio and smooth surface, giving $G$ an uncertainty of $\sim$10--20\%.\cite{Carrington1999} Our rectangular platelet samples with mirror-like surface fulfill these criteria. The sample is located on the axis of a solenoidal coil which has an AC field \textit{H}. The magntiude of \textit{H} is estimated to be $\sim$40 mOe. We report direct measurement of the in-plane penetration depth $\Delta$$\lambda$(\textit{T}) in this paper.     
\begin{figure}[h] 
	\centering \includegraphics[width=8cm,clip]{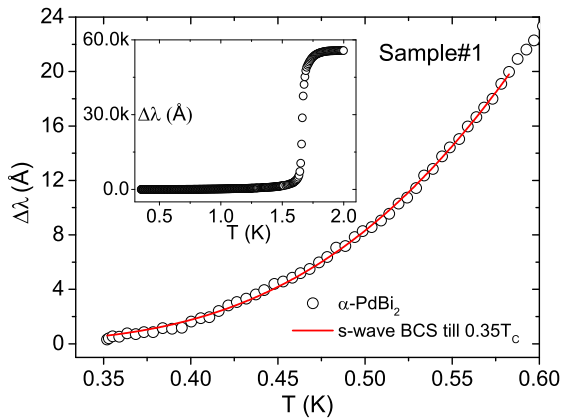}
	\caption{($\circ$) Low-temperature dependence of the in-plane penetration depth 
		$\Delta$$\lambda$(\textit{T}) in 
		$\alpha$-PdBi$_{2}$. The solid line is the fit to
		Eqn.~(\ref{eqn:lowTexponential}) from 0.35~K ($\sim$0.21$T_{c}$) to 0.58~K ($\sim$0.35$T_{c}$), with the fitting parameters, $\Delta(0)$/$k_{B}T_{c}$ = 2.00 and $\lambda(0)$ = 190~nm. Inset shows $\Delta$$\lambda$(\textit{T}) for the same sample over the full range.}
	\label{fig:Lambda}
\end{figure} 

Figure~\ref{fig:Lambda} shows $\Delta$$\lambda (T)$ in Sample\#1 of $\alpha$-PdBi$_{2}$ single crystal as a function of temperature up to 0.6~K. The inset shows $\Delta$$\lambda (T)$ for the sample plotted over the entire temperature range to temperatures above $T_{c}$ $\approx$ 1.66~K (onset of the superconducting transition). The 10\%-to-90\% transition width is only $\sim$0.03~K, showing that the measured crystal is of high quality. The low-temperature $\Delta$$\lambda (T)$ data is fitted to the standard \textit{s}-wave BCS model,\textcolor{red}{\cite{derivation59}}               
\begin{equation} \label{eqn:lowTexponential}
\Delta \lambda(T) \ = \lambda(0)\sqrt{\frac{\pi \Delta(0)}{2k_{B}T}}\exp\left(-\frac{\Delta(0)}{k_{B}T}\right),
\end{equation} 
with $\Delta(0)$ and $\lambda(0)$ as fitting parameters. As seen in Fig.~\ref{fig:Lambda}, the model fits our data well up to 0.35$T_{c}$ with the best fit obtained for $\Delta(0)$ = (2.00$\pm$0.02)$k_{B}T_{c}$ with $\lambda(0)$ = (190$\pm$10)~nm. The value of the obtained $\Delta(0)$/$k_{B}T_{c}$ is larger than the weak-coupling BCS value of 1.76, suggesting that $\alpha$-PdBi$_{2}$ is a moderate-coupling superconductor. 

In order to extract the in-plane normalized superfluid density $\rho_{s}(T)$ = [$\lambda^{2}$(0)/$\lambda^{2}(T)$] from $\Delta$$\lambda(T)$ data, we need to know the value of $\lambda(0)$. The previously-obtained value of $\lambda(0)$ is only an estimate, as it was obtained from fitting only low-temperature data.\cite{ProzorovlowT} In fitting $\rho_{s}(T)$ next we allow $\lambda(0)$ to be a fitting parameter. To calculate the theoretical $\rho_{s}(T)$, we used the expression for superfluid density for an isotropic \textit{s}-wave superconductor in the clean and local limits as shown below,\cite{Tinkham}
\begin{equation} \label{eqn:rhoth}
\rho_{s}(T) = 1 + 2\int^{\infty}_{0} \frac{\partial f}{\partial E}
d\varepsilon,
\end{equation}     
where $f$ = [exp(E/{\it k}$_{B}T$)+1]$^{-1}$ is the Fermi
function and $E = [\varepsilon^{2}$ + $\Delta (T)^{2}$]$^{1/2}$ is the Bogoliubov quasiparticle energy. The temperature dependence of the superconducting gap $\Delta(T)$ is given by\cite{Gross86}
\begin{equation} \label{eqn:gapinterpolate}
\Delta(\mathit{T})=\delta
_{sc}\mathit{kT}_{c}\tanh\left\{\frac{\pi}{\delta
	_{sc}}\sqrt{a\left(\frac{\Delta C}{C}\right)
	\left(\frac{T_{c}}{T}-1\right)}\right\},
\end{equation} where $\delta_{sc}$ = $\Delta(0)/k_{B}T_{c}$, a = 2/3 and $\Delta C/C
\equiv \Delta C/\gamma T_{c}$.

\begin{figure}[h]
	\centering \includegraphics[width=8cm,clip]{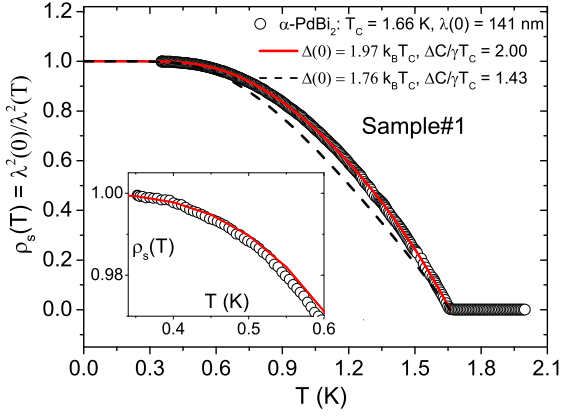}
	\caption{($\circ$) In-plane superfluid density $\rho_{s}(T)$ =
		[$\lambda^{2}$(0)/$\lambda^{2}(T)$] for $\alpha$-PdBi$_{2}$ Sample\#1
		calculated from $\Delta
		\lambda(T)$ data in Fig.~\ref{fig:Lambda} using $\lambda(0)$ = 141~nm. Solid line: Best fitted $\rho_{s}(T)$
		calculated from Eqn.~(\ref{eqn:rhoth}) using the parameters $\Delta(0)$/$k_{B}T_{c}$ = 1.97, $\Delta C/\gamma T_{c} =
		2.00$ and $T_{c}$ = 1.66~K. Dashed line: Calculated $\rho_{s}(T)$ using weak-coupling \textit{s}-wave parameters $\Delta(0)$/$k_{B}T_{c}$ = 1.76, $\Delta C/\gamma T_{c} =
		1.43$, for $T_{c}$ = 1.66~K. Inset shows $\rho_{s}(T)$ for the same sample up to 0.35$T_{c}$ along with the best fitting curve.} \label{fig:rhosample1}
\end{figure}

Keeping $T_{c}$ = 1.66~K fixed, and taking into account the $\sim$10\% uncertainty in the proportionality factor $G$,\cite{Carrington1999} we obtained the best fit with the following parameters: $\lambda(0)$ = (141$\pm$14)~nm, $\Delta(0)/k_{B}T_{c}$ = (1.97$\pm$0.04), and $\Delta C/\gamma T_{c}$ = (2.00$\pm$0.30), as shown as a solid line in Figure~\ref{fig:rhosample1}. The inset shows the low-temperature fit (up to 0.35$T_{c}$) between the experiment and theory for the same parameters. The dashed line in Figure~\ref{fig:rhosample1}, calculated using the BCS weak-coupling values of $\delta_{sc}$ = 1.76 and $\Delta C/\gamma T_{c}$ = 1.43, clearly does not fit the data. The fitted value of $\Delta(0)/k_{B}T_{c}$ agrees well with that obtained from the $\Delta$$\lambda(T)$ fit in Figure~\ref{fig:Lambda}, while the fitted value of $\lambda(0)$ agrees well with the value of 132~nm obtained for $\beta$-PdBi$_{2}$ from calorimetric studies.\cite{Samuely16}

To check the validity as well as the self-consistency of the obtained parameters, we use the strong-coupling equations \cite{Kresin75,Orlando79},
\begin{equation}
\eta _{\Delta }(\omega _{0})=1+5.3\left(\frac{T_{c}}{\omega _{0}}\right)^{2}\ln \left(\frac{%
\omega _{0}}{T_{c}}\right), \label{eqn:strongcouplingDelta}
\end{equation}

\begin{equation}
\eta _{Cv}(\omega _{0})=1+1.8\left(\frac{\pi T_{c}}{\omega _{0}}\right)^{2}\left(\ln (\frac{%
\omega _{0}}{T_{c}})+0.5\right), \label{eqn:strongcouplingCv}
\end{equation} where $\eta_{\Delta}$ and $\eta_{Cv}$ represent the correction factors that are required to be applied over the weak-coupling BCS gap ratio and specific heat jump, respectively, to get the corresponding values in the moderate-to-strong-coupling limits. Here $\omega_{0}$ is the characteristic (equivalent Einstein) frequency. If we use the fitting parameter $\Delta(0)/k_{B}T_{c}$ = 1.97, then we get the correction factor $\eta_{\Delta}$ = 1.97/1.76 = 1.12. Putting this value into Eqn.~(\ref{eqn:strongcouplingDelta}) with $T_{c}$ = 1.66~K gives $\omega_{0} \approx$ 16.9~K. Using this $\omega_{0}$ in Eqn.~(\ref{eqn:strongcouplingCv}) gives a specific heat jump of 2.08 --- this agrees well with the value of 2.00 obtained from the $\rho_{s}(T)$ fit and further supports our claim that $\alpha$-PdBi$_{2}$ is a moderately-coupled superconductor. 

In order to check the robustness and reproducibility of our data and analysis, we measured another single-crystal sample designated Sample\#2. The best fit of the superfluid density data, using the method described earlier, was obtained for the parameters $\lambda(0)$ = (134$\pm$13)~nm, $\Delta(0)/k_{B}T_{c}$ = (2.09$\pm$0.04), and $\Delta C/\gamma T_{c}$ = (2.10$\pm$0.29). We can see that (1) the fitted parameters of $\Delta(0)/k_{B}T_{c}$ and $\Delta C/\gamma T_{c}$ are consistent with each other via strong-coupling corrections, and (2) the obtained parameters for both $\alpha$-PdBi$_{2}$ samples are consistent with each other. 

\begin{figure}[h]
	\centering \includegraphics[width=8cm,clip]{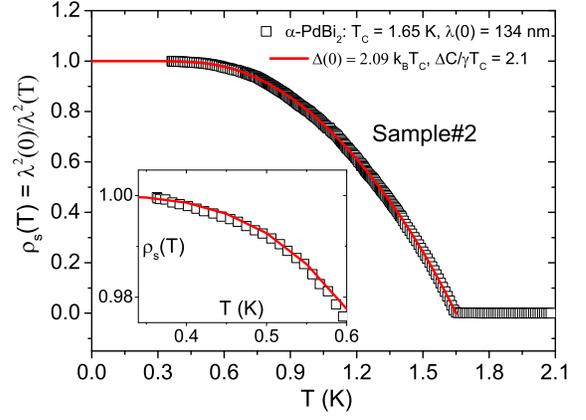}
	\caption{ ($\square$) In-plane superfluid density $\rho_{s}(T)$ =
		[$\lambda^{2}$(0)/$\lambda^{2}(T)$] for $\alpha$-PdBi$_{2}$ Sample\#2
		calculated from $\Delta
		\lambda(T)$ data in Fig.~\ref{fig:Lambda} using $\lambda(0)$ = 134~nm. Solid line: Best fitted $\rho_{s}(T)$
		calculated from Eqn.~(\ref{eqn:rhoth}) using the parameters $\Delta(0)$/$k_{B}T_{c}$ = 2.09, $\Delta C/\gamma T_{c} =
		2.10$ and $T_{c}$ = 1.65~K. Inset shows $\rho_{s}(T)$ for the same sample up to 0.35$T_{c}$ along with the best fitting curve.} 
	\label{fig:rhosample2}
\end{figure}

Thus, based on the analysis of the in-plane data in both the samples, we infer that $\alpha$-PdBi$_{2}$ is a single-gap isotropic moderately-coupled BCS superconductor with zero-temperature superconducting gap $\Delta(0)$/$k_{B}T_{c}$$\sim$2.0, and specific heat jump $\Delta C/\gamma T_{c}$$\sim$2.0, with superconducting transition temperature $T_{c}$$\sim$1.7~K. Measurements on the related centrosymmetric compound $\beta$-PdBi$_{2}$ using AC calorimetry, Hall-probe magnetometry and point-contact spectroscopic studies obtained $\Delta(0)$/$k_{B}T_{c}$ = 2.05 with $\Delta C/\gamma T_{c}$ $\approx$ 2.0.\cite{Samuely16,Liqiang16} Even in the NCS superconductor $\alpha$-PdBi, ultra-low-temperature scanning tunneling spectra \cite{Sun15} obtained a similar value of $\Delta(0)$/$k_{B}T_{c}$ $\approx$ 1.9. Clearly, these values are consistent with the parameters reported in our work.

In both $\beta$-PdBi$_{2}$ and $\alpha$-PdBi, even though multiple theoretical calculations as well as experimental observations have clearly pointed out the existence of topological states, the bulk superconducting ground state always seems to be topologically trivial, consistent with a BCS-like \textit{s}-wave order parameter. Our electronic structure calculations show a metallic normal-state of $\alpha$-PdBi$_{2}$, similar to $\alpha$-PdBi and $\beta$-PdBi$_{2}$.\cite{Zhu17} It leads us to believe that the three compounds have the same nature of the superconducting state. It has been suggested that for Type-II superconductors, the surface Andreev bound states consisting of Majorana Fermions are expected to decay into the bulk within a few coherence lengths $\xi$. Using the value of $\xi$$\approx$20~nm for $\beta$-PdBi$_{2}$ from calorimetric measurements,\cite{Samuely16}, and $\xi$$\approx$66~nm for $\alpha$-PdBi from STS measurements,\cite{Sun15} these states should have a spatial extent of $\sim$100~nm from the surface. Given our fitted value of $\lambda(0)$$\approx$140~nm, our penetration depth measurements is able to measure field penetration from $\sim$140~nm \textit{inwards}, with Angstrom resolution. To elaborate on this, even at zero temperature, the magnetic field has already penetrated through the sample over a distance of $\sim$140~nm in our sample. Hence any gapless excitation, which exists over the aforementioned length scale of $\sim$100 nm from the surface, will not be detected by our technique. This implies that we are barely able to probe the topological surface states in $\alpha$-PdBi$_{2}$ and thus the absence of a low-temperature power law in our data does not necessarily rule out the presence of surface states in this material. More surface-sensitive spectroscopic measurements such as point-contact Andreev spectroscopy and ARPES should give direct evidence of topologically-protected surface states in this class of possible stiochiometric TSCs. Additionally, $\mu$SR and calorimetric measurements should be performed to validate the parameters we have reported in this paper.               

In conclusion, we report measurements of the in-plane London penetration depth $\lambda$ in single crystals of $\alpha$-PdBi$_{2}$ down to 0.35~K. Fits to the measured penetration depth $\Delta \lambda $(\textit{T}) and the normalized superfluid density $\rho_{s}$(\textit{T}) suggest the existence of a moderate-coupling single \textit{s}-wave gap in this material. Comparison with the other related compounds shows that the superconducting order parameter has a similar pairing symmetry across the Pd-Bi family of superconductors. Data from our high-resolution and surface-sensitive penetration depth setup did not observe gapless excitations on the surface, thus could not detect any signature of the topological nature of superconductivity in this material. Further measurements from other surface-sensitive experimental techniques have to be performed to ascertain the presence or absence of topologically-protected surface states in $\alpha$-PdBi$_{2}$.      

We acknowledge funding from the Ministry of Education (MOE) Tier 1 Grant (RG13/12 and MOE2015-T2-2-065). The work at Tokyo Tech. was supported by a CREST project from Japan Science and Technology Agency (JST) and  a Grants-in-Aid for Scientific Research (B) from Japan Society for the Promotion of Science (JSPS). This work was supported, in part, by the Center for Integrated Nanotechnologies, a U.S. DOE Office of Basic Energy Sciences user facility.

\bibliography{AlphaPdBi2}

\begin{thebibliography}{43}%
\makeatletter
\providecommand \@ifxundefined [1]{%
 \@ifx{#1\undefined}
}%
\providecommand \@ifnum [1]{%
 \ifnum #1\expandafter \@firstoftwo
 \else \expandafter \@secondoftwo
 \fi
}%
\providecommand \@ifx [1]{%
 \ifx #1\expandafter \@firstoftwo
 \else \expandafter \@secondoftwo
 \fi
}%
\providecommand \natexlab [1]{#1}%
\providecommand \enquote  [1]{``#1''}%
\providecommand \bibnamefont  [1]{#1}%
\providecommand \bibfnamefont [1]{#1}%
\providecommand \citenamefont [1]{#1}%
\providecommand \href@noop [0]{\@secondoftwo}%
\providecommand \href [0]{\begingroup \@sanitize@url \@href}%
\providecommand \@href[1]{\@@startlink{#1}\@@href}%
\providecommand \@@href[1]{\endgroup#1\@@endlink}%
\providecommand \@sanitize@url [0]{\catcode `\\12\catcode `\$12\catcode
  `\&12\catcode `\#12\catcode `\^12\catcode `\_12\catcode `\%12\relax}%
\providecommand \@@startlink[1]{}%
\providecommand \@@endlink[0]{}%
\providecommand \url  [0]{\begingroup\@sanitize@url \@url }%
\providecommand \@url [1]{\endgroup\@href {#1}{\urlprefix }}%
\providecommand \urlprefix  [0]{URL }%
\providecommand \Eprint [0]{\href }%
\providecommand \doibase [0]{http://dx.doi.org/}%
\providecommand \selectlanguage [0]{\@gobble}%
\providecommand \bibinfo  [0]{\@secondoftwo}%
\providecommand \bibfield  [0]{\@secondoftwo}%
\providecommand \translation [1]{[#1]}%
\providecommand \BibitemOpen [0]{}%
\providecommand \bibitemStop [0]{}%
\providecommand \bibitemNoStop [0]{.\EOS\space}%
\providecommand \EOS [0]{\spacefactor3000\relax}%
\providecommand \BibitemShut  [1]{\csname bibitem#1\endcsname}%
\let\auto@bib@innerbib\@empty
%</preamble>
\bibitem [{\citenamefont {Sakano}\ \emph {et~al.}(2015)\citenamefont {Sakano},
  \citenamefont {Okawa}, \citenamefont {Kanou}, \citenamefont {Sanjo},
  \citenamefont {Okuda}, \citenamefont {Sasagawa},\ and\ \citenamefont
  {Ishizaka}}]{Sasagawa15}%
  \BibitemOpen
  \bibfield  {author} {\bibinfo {author} {\bibfnamefont {M.}~\bibnamefont
  {Sakano}}, \bibinfo {author} {\bibfnamefont {K.}~\bibnamefont {Okawa}},
  \bibinfo {author} {\bibfnamefont {M.}~\bibnamefont {Kanou}}, \bibinfo
  {author} {\bibfnamefont {H.}~\bibnamefont {Sanjo}}, \bibinfo {author}
  {\bibfnamefont {T.}~\bibnamefont {Okuda}}, \bibinfo {author} {\bibfnamefont
  {T.}~\bibnamefont {Sasagawa}}, \ and\ \bibinfo {author} {\bibfnamefont
  {K.}~\bibnamefont {Ishizaka}},\ }\href@noop {} {\bibfield  {journal}
  {\bibinfo  {journal} {Nat. Commun.}\ }\textbf {\bibinfo {volume} {6}},\
  \bibinfo {pages} {8595} (\bibinfo {year} {2015})}\BibitemShut {NoStop}%
\bibitem [{\citenamefont {Che}\ \emph {et~al.}(2016)\citenamefont {Che},
  \citenamefont {Le}, \citenamefont {Xu}, \citenamefont {Xing}, \citenamefont
  {Shi}, \citenamefont {Xu},\ and\ \citenamefont {Lu}}]{Liqiang16}%
  \BibitemOpen
  \bibfield  {author} {\bibinfo {author} {\bibfnamefont {L.}~\bibnamefont
  {Che}}, \bibinfo {author} {\bibfnamefont {T.}~\bibnamefont {Le}}, \bibinfo
  {author} {\bibfnamefont {C.~Q.}\ \bibnamefont {Xu}}, \bibinfo {author}
  {\bibfnamefont {X.~Z.}\ \bibnamefont {Xing}}, \bibinfo {author}
  {\bibfnamefont {Z.}~\bibnamefont {Shi}}, \bibinfo {author} {\bibfnamefont
  {X.}~\bibnamefont {Xu}}, \ and\ \bibinfo {author} {\bibfnamefont
  {X.}~\bibnamefont {Lu}},\ }\href@noop {} {\bibfield  {journal} {\bibinfo
  {journal} {Phys. Rev. B}\ }\textbf {\bibinfo {volume} {94}},\ \bibinfo
  {pages} {024519} (\bibinfo {year} {2016})}\BibitemShut {NoStop}%
\bibitem [{\citenamefont {Imai}\ \emph {et~al.}(2012)\citenamefont {Imai},
  \citenamefont {Nabeshima}, \citenamefont {Yoshinaka}, \citenamefont
  {Miyatani}, \citenamefont {Kondo}, \citenamefont {Komiya}, \citenamefont
  {Tsukada},\ and\ \citenamefont {Maeda}}]{Imai12}%
  \BibitemOpen
  \bibfield  {author} {\bibinfo {author} {\bibfnamefont {Y.}~\bibnamefont
  {Imai}}, \bibinfo {author} {\bibfnamefont {F.}~\bibnamefont {Nabeshima}},
  \bibinfo {author} {\bibfnamefont {T.}~\bibnamefont {Yoshinaka}}, \bibinfo
  {author} {\bibfnamefont {K.}~\bibnamefont {Miyatani}}, \bibinfo {author}
  {\bibfnamefont {R.}~\bibnamefont {Kondo}}, \bibinfo {author} {\bibfnamefont
  {S.}~\bibnamefont {Komiya}}, \bibinfo {author} {\bibfnamefont
  {I.}~\bibnamefont {Tsukada}}, \ and\ \bibinfo {author} {\bibfnamefont
  {A.}~\bibnamefont {Maeda}},\ }\href@noop {} {\bibfield  {journal} {\bibinfo
  {journal} {J. Phys. Soc. Jpn.}\ }\textbf {\bibinfo {volume} {81}},\ \bibinfo
  {pages} {113708} (\bibinfo {year} {2012})}\BibitemShut {NoStop}%
\bibitem [{\citenamefont {Herrera}\ \emph {et~al.}(2015)\citenamefont
  {Herrera}, \citenamefont {Guillam\'on}, \citenamefont {Galvis}, \citenamefont
  {Correa}, \citenamefont {Fente}, \citenamefont {Luccas}, \citenamefont
  {Mompean}, \citenamefont {Garc\'{\i}a-Hern\'andez}, \citenamefont {Vieira},
  \citenamefont {Brison},\ and\ \citenamefont {Suderow}}]{Herrera15}%
  \BibitemOpen
  \bibfield  {author} {\bibinfo {author} {\bibfnamefont {E.}~\bibnamefont
  {Herrera}}, \bibinfo {author} {\bibfnamefont {I.}~\bibnamefont
  {Guillam\'on}}, \bibinfo {author} {\bibfnamefont {J.~A.}\ \bibnamefont
  {Galvis}}, \bibinfo {author} {\bibfnamefont {A.}~\bibnamefont {Correa}},
  \bibinfo {author} {\bibfnamefont {A.}~\bibnamefont {Fente}}, \bibinfo
  {author} {\bibfnamefont {R.~F.}\ \bibnamefont {Luccas}}, \bibinfo {author}
  {\bibfnamefont {F.~J.}\ \bibnamefont {Mompean}}, \bibinfo {author}
  {\bibfnamefont {M.}~\bibnamefont {Garc\'{\i}a-Hern\'andez}}, \bibinfo
  {author} {\bibfnamefont {S.}~\bibnamefont {Vieira}}, \bibinfo {author}
  {\bibfnamefont {J.~P.}\ \bibnamefont {Brison}}, \ and\ \bibinfo {author}
  {\bibfnamefont {H.}~\bibnamefont {Suderow}},\ }\href@noop {} {\bibfield
  {journal} {\bibinfo  {journal} {Phys. Rev. B}\ }\textbf {\bibinfo {volume}
  {92}},\ \bibinfo {pages} {054507} (\bibinfo {year} {2015})}\BibitemShut
  {NoStop}%
\bibitem [{\citenamefont {Biswas}\ \emph {et~al.}(2016)\citenamefont {Biswas},
  \citenamefont {Mazzone}, \citenamefont {Sibille}, \citenamefont
  {Pomjakushina}, \citenamefont {Conder}, \citenamefont {Luetkens},
  \citenamefont {Baines}, \citenamefont {Gavilano}, \citenamefont {Kenzelmann},
  \citenamefont {Amato},\ and\ \citenamefont {Morenzoni}}]{Biswas16}%
  \BibitemOpen
  \bibfield  {author} {\bibinfo {author} {\bibfnamefont {P.~K.}\ \bibnamefont
  {Biswas}}, \bibinfo {author} {\bibfnamefont {D.~G.}\ \bibnamefont {Mazzone}},
  \bibinfo {author} {\bibfnamefont {R.}~\bibnamefont {Sibille}}, \bibinfo
  {author} {\bibfnamefont {E.}~\bibnamefont {Pomjakushina}}, \bibinfo {author}
  {\bibfnamefont {K.}~\bibnamefont {Conder}}, \bibinfo {author} {\bibfnamefont
  {H.}~\bibnamefont {Luetkens}}, \bibinfo {author} {\bibfnamefont
  {C.}~\bibnamefont {Baines}}, \bibinfo {author} {\bibfnamefont {J.~L.}\
  \bibnamefont {Gavilano}}, \bibinfo {author} {\bibfnamefont {M.}~\bibnamefont
  {Kenzelmann}}, \bibinfo {author} {\bibfnamefont {A.}~\bibnamefont {Amato}}, \
  and\ \bibinfo {author} {\bibfnamefont {E.}~\bibnamefont {Morenzoni}},\
  }\href@noop {} {\bibfield  {journal} {\bibinfo  {journal} {Phys. Rev. B}\
  }\textbf {\bibinfo {volume} {93}},\ \bibinfo {pages} {220504(R)} (\bibinfo
  {year} {2016})}\BibitemShut {NoStop}%
\bibitem [{\citenamefont {Ka\ifmmode \check{c}\else
  \v{c}\fi{}mar\ifmmode~\check{c}\else \v{c}\fi{}\'{\i}k}\ \emph
  {et~al.}(2016)\citenamefont {Ka\ifmmode \check{c}\else
  \v{c}\fi{}mar\ifmmode~\check{c}\else \v{c}\fi{}\'{\i}k}, \citenamefont
  {Pribulov\'a}, \citenamefont {Samuely}, \citenamefont {Szab\'o},
  \citenamefont {Cambel}, \citenamefont {\ifmmode~\check{S}\else
  \v{S}\fi{}olt\'ys}, \citenamefont {Herrera}, \citenamefont {Suderow},
  \citenamefont {Correa-Orellana}, \citenamefont {Prabhakaran},\ and\
  \citenamefont {Samuely}}]{Samuely16}%
  \BibitemOpen
  \bibfield  {author} {\bibinfo {author} {\bibfnamefont {J.}~\bibnamefont
  {Ka\ifmmode \check{c}\else \v{c}\fi{}mar\ifmmode~\check{c}\else
  \v{c}\fi{}\'{\i}k}}, \bibinfo {author} {\bibfnamefont {Z.}~\bibnamefont
  {Pribulov\'a}}, \bibinfo {author} {\bibfnamefont {T.}~\bibnamefont
  {Samuely}}, \bibinfo {author} {\bibfnamefont {P.}~\bibnamefont {Szab\'o}},
  \bibinfo {author} {\bibfnamefont {V.}~\bibnamefont {Cambel}}, \bibinfo
  {author} {\bibfnamefont {J.}~\bibnamefont {\ifmmode~\check{S}\else
  \v{S}\fi{}olt\'ys}}, \bibinfo {author} {\bibfnamefont {E.}~\bibnamefont
  {Herrera}}, \bibinfo {author} {\bibfnamefont {H.}~\bibnamefont {Suderow}},
  \bibinfo {author} {\bibfnamefont {A.}~\bibnamefont {Correa-Orellana}},
  \bibinfo {author} {\bibfnamefont {D.}~\bibnamefont {Prabhakaran}}, \ and\
  \bibinfo {author} {\bibfnamefont {P.}~\bibnamefont {Samuely}},\ }\href@noop
  {} {\bibfield  {journal} {\bibinfo  {journal} {Phys. Rev. B}\ }\textbf
  {\bibinfo {volume} {93}},\ \bibinfo {pages} {144502} (\bibinfo {year}
  {2016})}\BibitemShut {NoStop}%
\bibitem [{\citenamefont {Zhuravlev}(1957)}]{Zhuravlev57}%
  \BibitemOpen
  \bibfield  {author} {\bibinfo {author} {\bibfnamefont {N.~N.}\ \bibnamefont
  {Zhuravlev}},\ }\href@noop {} {\bibfield  {journal} {\bibinfo  {journal} {Zh.
  Eksp. Teor. Fiz.}\ }\textbf {\bibinfo {volume} {32}},\ \bibinfo {pages}
  {1305} (\bibinfo {year} {1957})}\BibitemShut {NoStop}%
\bibitem [{\citenamefont {Bhatt}\ and\ \citenamefont
  {Schubert}(1979)}]{Bhatt79}%
  \BibitemOpen
  \bibfield  {author} {\bibinfo {author} {\bibfnamefont {Y.}~\bibnamefont
  {Bhatt}}\ and\ \bibinfo {author} {\bibfnamefont {K.~J.}\ \bibnamefont
  {Schubert}},\ }\href@noop {} {\bibfield  {journal} {\bibinfo  {journal} {J.
  Less-Common Met.}\ }\textbf {\bibinfo {volume} {64}},\ \bibinfo {pages} {17}
  (\bibinfo {year} {1979})}\BibitemShut {NoStop}%
\bibitem [{\citenamefont {Joshi}\ \emph {et~al.}(2011)\citenamefont {Joshi},
  \citenamefont {Thamizhavel},\ and\ \citenamefont {Ramakrishnan}}]{Joshi11}%
  \BibitemOpen
  \bibfield  {author} {\bibinfo {author} {\bibfnamefont {B.}~\bibnamefont
  {Joshi}}, \bibinfo {author} {\bibfnamefont {A.}~\bibnamefont {Thamizhavel}},
  \ and\ \bibinfo {author} {\bibfnamefont {S.}~\bibnamefont {Ramakrishnan}},\
  }\href@noop {} {\bibfield  {journal} {\bibinfo  {journal} {Phys. Rev. B}\
  }\textbf {\bibinfo {volume} {84}},\ \bibinfo {pages} {064518} (\bibinfo
  {year} {2011})}\BibitemShut {NoStop}%
\bibitem [{\citenamefont {Gor'kov}\ and\ \citenamefont
  {Rashba}(2001)}]{Rashba01}%
  \BibitemOpen
  \bibfield  {author} {\bibinfo {author} {\bibfnamefont {L.~P.}\ \bibnamefont
  {Gor'kov}}\ and\ \bibinfo {author} {\bibfnamefont {E.~I.}\ \bibnamefont
  {Rashba}},\ }\href@noop {} {\bibfield  {journal} {\bibinfo  {journal} {Phys.
  Rev. Lett.}\ }\textbf {\bibinfo {volume} {87}},\ \bibinfo {pages} {037004}
  (\bibinfo {year} {2001})}\BibitemShut {NoStop}%
\bibitem [{\citenamefont {Jiao}\ \emph {et~al.}(2014)\citenamefont {Jiao},
  \citenamefont {Zhang}, \citenamefont {Chen}, \citenamefont {Weng},
  \citenamefont {Shao}, \citenamefont {Feng}, \citenamefont {Lu}, \citenamefont
  {Joshi}, \citenamefont {Thamizhavel}, \citenamefont {Ramakrishnan},\ and\
  \citenamefont {Yuan}}]{Jiao14}%
  \BibitemOpen
  \bibfield  {author} {\bibinfo {author} {\bibfnamefont {L.}~\bibnamefont
  {Jiao}}, \bibinfo {author} {\bibfnamefont {J.~L.}\ \bibnamefont {Zhang}},
  \bibinfo {author} {\bibfnamefont {Y.}~\bibnamefont {Chen}}, \bibinfo {author}
  {\bibfnamefont {Z.~F.}\ \bibnamefont {Weng}}, \bibinfo {author}
  {\bibfnamefont {Y.~M.}\ \bibnamefont {Shao}}, \bibinfo {author}
  {\bibfnamefont {J.~Y.}\ \bibnamefont {Feng}}, \bibinfo {author}
  {\bibfnamefont {X.}~\bibnamefont {Lu}}, \bibinfo {author} {\bibfnamefont
  {B.}~\bibnamefont {Joshi}}, \bibinfo {author} {\bibfnamefont
  {A.}~\bibnamefont {Thamizhavel}}, \bibinfo {author} {\bibfnamefont
  {S.}~\bibnamefont {Ramakrishnan}}, \ and\ \bibinfo {author} {\bibfnamefont
  {H.~Q.}\ \bibnamefont {Yuan}},\ }\href@noop {} {\bibfield  {journal}
  {\bibinfo  {journal} {Phys. Rev. B}\ }\textbf {\bibinfo {volume} {89}},\
  \bibinfo {pages} {060507} (\bibinfo {year} {2014})}\BibitemShut {NoStop}%
\bibitem [{\citenamefont {Smidman}\ \emph {et~al.}(2016)\citenamefont
  {Smidman}, \citenamefont {Salamon}, \citenamefont {Yuan},\ and\ \citenamefont
  {Agterberg}}]{Smidman16}%
  \BibitemOpen
  \bibfield  {author} {\bibinfo {author} {\bibfnamefont {M.}~\bibnamefont
  {Smidman}}, \bibinfo {author} {\bibfnamefont {M.~B.}\ \bibnamefont
  {Salamon}}, \bibinfo {author} {\bibfnamefont {H.~Q.}\ \bibnamefont {Yuan}}, \
  and\ \bibinfo {author} {\bibfnamefont {D.~F.}\ \bibnamefont {Agterberg}},\
  }\href@noop {} {\bibfield  {journal} {\bibinfo  {journal} {arXiv:1609.05953
  [cond-mat.supr-con]}\ } (\bibinfo {year} {2016})}\BibitemShut {NoStop}%
\bibitem [{\citenamefont {Mackenzie}\ \emph {et~al.}(1998)\citenamefont
  {Mackenzie}, \citenamefont {Haselwimmer}, \citenamefont {Tyler},
  \citenamefont {Lonzarich}, \citenamefont {Mori}, \citenamefont {Nishizaki},\
  and\ \citenamefont {Maeno}}]{Spinttriplet98}%
  \BibitemOpen
  \bibfield  {author} {\bibinfo {author} {\bibfnamefont {A.~P.}\ \bibnamefont
  {Mackenzie}}, \bibinfo {author} {\bibfnamefont {R.~K.~W.}\ \bibnamefont
  {Haselwimmer}}, \bibinfo {author} {\bibfnamefont {A.~W.}\ \bibnamefont
  {Tyler}}, \bibinfo {author} {\bibfnamefont {G.~G.}\ \bibnamefont
  {Lonzarich}}, \bibinfo {author} {\bibfnamefont {Y.}~\bibnamefont {Mori}},
  \bibinfo {author} {\bibfnamefont {S.}~\bibnamefont {Nishizaki}}, \ and\
  \bibinfo {author} {\bibfnamefont {Y.}~\bibnamefont {Maeno}},\ }\href@noop {}
  {\bibfield  {journal} {\bibinfo  {journal} {Phys. Rev. Lett.}\ }\textbf
  {\bibinfo {volume} {80}},\ \bibinfo {pages} {161} (\bibinfo {year}
  {1998})}\BibitemShut {NoStop}%
\bibitem [{\citenamefont {Brun}\ \emph {et~al.}(2014)\citenamefont {Brun},
  \citenamefont {Cren}, \citenamefont {Cherkez}, \citenamefont {Debontridder},
  \citenamefont {Pons}, \citenamefont {Fokin}, \citenamefont {Tringides},
  \citenamefont {Bozhko}, \citenamefont {Ioffe}, \citenamefont {Altshuler},\
  and\ \citenamefont {Roditchev}}]{Spinttriplet14}%
  \BibitemOpen
  \bibfield  {author} {\bibinfo {author} {\bibfnamefont {C.}~\bibnamefont
  {Brun}}, \bibinfo {author} {\bibfnamefont {T.}~\bibnamefont {Cren}}, \bibinfo
  {author} {\bibfnamefont {V.}~\bibnamefont {Cherkez}}, \bibinfo {author}
  {\bibfnamefont {F.}~\bibnamefont {Debontridder}}, \bibinfo {author}
  {\bibfnamefont {S.}~\bibnamefont {Pons}}, \bibinfo {author} {\bibfnamefont
  {D.}~\bibnamefont {Fokin}}, \bibinfo {author} {\bibfnamefont {M.~C.}\
  \bibnamefont {Tringides}}, \bibinfo {author} {\bibfnamefont {S.}~\bibnamefont
  {Bozhko}}, \bibinfo {author} {\bibfnamefont {L.~B.}\ \bibnamefont {Ioffe}},
  \bibinfo {author} {\bibfnamefont {B.~L.}\ \bibnamefont {Altshuler}}, \ and\
  \bibinfo {author} {\bibfnamefont {D.}~\bibnamefont {Roditchev}},\ }\href@noop
  {} {\bibfield  {journal} {\bibinfo  {journal} {Nat Phys}\ }\textbf {\bibinfo
  {volume} {10}},\ \bibinfo {pages} {444} (\bibinfo {year} {2014})}\BibitemShut
  {NoStop}%
\bibitem [{\citenamefont {Neupane}\ \emph {et~al.}(2016)\citenamefont
  {Neupane}, \citenamefont {Alidoust}, \citenamefont {Hosen}, \citenamefont
  {Zhu}, \citenamefont {Dimitri}, \citenamefont {Xu}, \citenamefont {Dhakal},
  \citenamefont {Sankar}, \citenamefont {Belopolski}, \citenamefont {Sanchez},
  \citenamefont {Chang}, \citenamefont {Jeng}, \citenamefont {Miyamoto},
  \citenamefont {Okuda}, \citenamefont {Lin}, \citenamefont {Bansil},
  \citenamefont {Kaczorowski}, \citenamefont {Chou}, \citenamefont {Hasan},\
  and\ \citenamefont {Durakiewicz}}]{Neupane16}%
  \BibitemOpen
  \bibfield  {author} {\bibinfo {author} {\bibfnamefont {M.}~\bibnamefont
  {Neupane}}, \bibinfo {author} {\bibfnamefont {N.}~\bibnamefont {Alidoust}},
  \bibinfo {author} {\bibfnamefont {M.~M.}\ \bibnamefont {Hosen}}, \bibinfo
  {author} {\bibfnamefont {J.-X.}\ \bibnamefont {Zhu}}, \bibinfo {author}
  {\bibfnamefont {K.}~\bibnamefont {Dimitri}}, \bibinfo {author} {\bibfnamefont
  {S.-Y.}\ \bibnamefont {Xu}}, \bibinfo {author} {\bibfnamefont
  {N.}~\bibnamefont {Dhakal}}, \bibinfo {author} {\bibfnamefont
  {R.}~\bibnamefont {Sankar}}, \bibinfo {author} {\bibfnamefont
  {I.}~\bibnamefont {Belopolski}}, \bibinfo {author} {\bibfnamefont {D.~S.}\
  \bibnamefont {Sanchez}}, \bibinfo {author} {\bibfnamefont {T.-R.}\
  \bibnamefont {Chang}}, \bibinfo {author} {\bibfnamefont {H.-T.}\ \bibnamefont
  {Jeng}}, \bibinfo {author} {\bibfnamefont {K.}~\bibnamefont {Miyamoto}},
  \bibinfo {author} {\bibfnamefont {T.}~\bibnamefont {Okuda}}, \bibinfo
  {author} {\bibfnamefont {H.}~\bibnamefont {Lin}}, \bibinfo {author}
  {\bibfnamefont {A.}~\bibnamefont {Bansil}}, \bibinfo {author} {\bibfnamefont
  {D.}~\bibnamefont {Kaczorowski}}, \bibinfo {author} {\bibfnamefont
  {F.}~\bibnamefont {Chou}}, \bibinfo {author} {\bibfnamefont {M.~Z.}\
  \bibnamefont {Hasan}}, \ and\ \bibinfo {author} {\bibfnamefont
  {T.}~\bibnamefont {Durakiewicz}},\ }\href@noop {} {\bibfield  {journal}
  {\bibinfo  {journal} {Nat. Commun.}\ }\textbf {\bibinfo {volume} {7}},\
  \bibinfo {pages} {13315} (\bibinfo {year} {2016})}\BibitemShut {NoStop}%
\bibitem [{\citenamefont {Sun}\ \emph {et~al.}(2015)\citenamefont {Sun},
  \citenamefont {Enayat}, \citenamefont {Maldonado}, \citenamefont {Lithgow},
  \citenamefont {Yelland}, \citenamefont {Peets}, \citenamefont {Yaresko},
  \citenamefont {Schnyder},\ and\ \citenamefont {Wahl}}]{Sun15}%
  \BibitemOpen
  \bibfield  {author} {\bibinfo {author} {\bibfnamefont {Z.}~\bibnamefont
  {Sun}}, \bibinfo {author} {\bibfnamefont {M.}~\bibnamefont {Enayat}},
  \bibinfo {author} {\bibfnamefont {A.}~\bibnamefont {Maldonado}}, \bibinfo
  {author} {\bibfnamefont {C.}~\bibnamefont {Lithgow}}, \bibinfo {author}
  {\bibfnamefont {E.}~\bibnamefont {Yelland}}, \bibinfo {author} {\bibfnamefont
  {D.~C.}\ \bibnamefont {Peets}}, \bibinfo {author} {\bibfnamefont
  {A.}~\bibnamefont {Yaresko}}, \bibinfo {author} {\bibfnamefont {A.~P.}\
  \bibnamefont {Schnyder}}, \ and\ \bibinfo {author} {\bibfnamefont
  {P.}~\bibnamefont {Wahl}},\ }\href@noop {} {\bibfield  {journal} {\bibinfo
  {journal} {Nat. Commun.}\ }\textbf {\bibinfo {volume} {6}},\ \bibinfo {pages}
  {6633} (\bibinfo {year} {2015})}\BibitemShut {NoStop}%
\bibitem [{\citenamefont {Hor}\ \emph {et~al.}(2010)\citenamefont {Hor},
  \citenamefont {Williams}, \citenamefont {Checkelsky}, \citenamefont
  {Roushan}, \citenamefont {Seo}, \citenamefont {Xu}, \citenamefont
  {Zandbergen}, \citenamefont {Yazdani}, \citenamefont {Ong},\ and\
  \citenamefont {Cava}}]{Hor10}%
  \BibitemOpen
  \bibfield  {author} {\bibinfo {author} {\bibfnamefont {Y.~S.}\ \bibnamefont
  {Hor}}, \bibinfo {author} {\bibfnamefont {A.~J.}\ \bibnamefont {Williams}},
  \bibinfo {author} {\bibfnamefont {J.~G.}\ \bibnamefont {Checkelsky}},
  \bibinfo {author} {\bibfnamefont {P.}~\bibnamefont {Roushan}}, \bibinfo
  {author} {\bibfnamefont {J.}~\bibnamefont {Seo}}, \bibinfo {author}
  {\bibfnamefont {Q.}~\bibnamefont {Xu}}, \bibinfo {author} {\bibfnamefont
  {H.~W.}\ \bibnamefont {Zandbergen}}, \bibinfo {author} {\bibfnamefont
  {A.}~\bibnamefont {Yazdani}}, \bibinfo {author} {\bibfnamefont {N.~P.}\
  \bibnamefont {Ong}}, \ and\ \bibinfo {author} {\bibfnamefont {R.~J.}\
  \bibnamefont {Cava}},\ }\href@noop {} {\bibfield  {journal} {\bibinfo
  {journal} {Phys. Rev. Lett.}\ }\textbf {\bibinfo {volume} {104}},\ \bibinfo
  {pages} {057001} (\bibinfo {year} {2010})}\BibitemShut {NoStop}%
\bibitem [{\citenamefont {Kriener}\ \emph {et~al.}(2011)\citenamefont
  {Kriener}, \citenamefont {Segawa}, \citenamefont {Ren}, \citenamefont
  {Sasaki},\ and\ \citenamefont {Ando}}]{Kriener11}%
  \BibitemOpen
  \bibfield  {author} {\bibinfo {author} {\bibfnamefont {M.}~\bibnamefont
  {Kriener}}, \bibinfo {author} {\bibfnamefont {K.}~\bibnamefont {Segawa}},
  \bibinfo {author} {\bibfnamefont {Z.}~\bibnamefont {Ren}}, \bibinfo {author}
  {\bibfnamefont {S.}~\bibnamefont {Sasaki}}, \ and\ \bibinfo {author}
  {\bibfnamefont {Y.}~\bibnamefont {Ando}},\ }\href@noop {} {\bibfield
  {journal} {\bibinfo  {journal} {Phys. Rev. Lett.}\ }\textbf {\bibinfo
  {volume} {106}},\ \bibinfo {pages} {127004} (\bibinfo {year}
  {2011})}\BibitemShut {NoStop}%
\bibitem [{\citenamefont {Smylie}\ \emph {et~al.}(2016)\citenamefont {Smylie},
  \citenamefont {Claus}, \citenamefont {Welp}, \citenamefont {Kwok},
  \citenamefont {Qiu}, \citenamefont {Hor},\ and\ \citenamefont
  {Snezhko}}]{Smylie16}%
  \BibitemOpen
  \bibfield  {author} {\bibinfo {author} {\bibfnamefont {M.~P.}\ \bibnamefont
  {Smylie}}, \bibinfo {author} {\bibfnamefont {H.}~\bibnamefont {Claus}},
  \bibinfo {author} {\bibfnamefont {U.}~\bibnamefont {Welp}}, \bibinfo {author}
  {\bibfnamefont {W.-K.}\ \bibnamefont {Kwok}}, \bibinfo {author}
  {\bibfnamefont {Y.}~\bibnamefont {Qiu}}, \bibinfo {author} {\bibfnamefont
  {Y.~S.}\ \bibnamefont {Hor}}, \ and\ \bibinfo {author} {\bibfnamefont
  {A.}~\bibnamefont {Snezhko}},\ }\href@noop {} {\bibfield  {journal} {\bibinfo
   {journal} {Phys. Rev. B}\ }\textbf {\bibinfo {volume} {94}},\ \bibinfo
  {pages} {180510} (\bibinfo {year} {2016})}\BibitemShut {NoStop}%
\bibitem [{\citenamefont {Sasaki}\ \emph {et~al.}(2012)\citenamefont {Sasaki},
  \citenamefont {Ren}, \citenamefont {Taskin}, \citenamefont {Segawa},
  \citenamefont {Fu},\ and\ \citenamefont {Ando}}]{Sasaki12}%
  \BibitemOpen
  \bibfield  {author} {\bibinfo {author} {\bibfnamefont {S.}~\bibnamefont
  {Sasaki}}, \bibinfo {author} {\bibfnamefont {Z.}~\bibnamefont {Ren}},
  \bibinfo {author} {\bibfnamefont {A.~A.}\ \bibnamefont {Taskin}}, \bibinfo
  {author} {\bibfnamefont {K.}~\bibnamefont {Segawa}}, \bibinfo {author}
  {\bibfnamefont {L.}~\bibnamefont {Fu}}, \ and\ \bibinfo {author}
  {\bibfnamefont {Y.}~\bibnamefont {Ando}},\ }\href@noop {} {\bibfield
  {journal} {\bibinfo  {journal} {Phys. Rev. Lett.}\ }\textbf {\bibinfo
  {volume} {109}},\ \bibinfo {pages} {217004} (\bibinfo {year}
  {2012})}\BibitemShut {NoStop}%
\bibitem [{\citenamefont {Guan}\ \emph {et~al.}(2016)\citenamefont {Guan},
  \citenamefont {Chen}, \citenamefont {Chu}, \citenamefont {Sankar},
  \citenamefont {Chou}, \citenamefont {Jeng}, \citenamefont {Chang},\ and\
  \citenamefont {Chuang}}]{Guan16}%
  \BibitemOpen
  \bibfield  {author} {\bibinfo {author} {\bibfnamefont {S.~Y.}\ \bibnamefont
  {Guan}}, \bibinfo {author} {\bibfnamefont {P.~J.}\ \bibnamefont {Chen}},
  \bibinfo {author} {\bibfnamefont {M.~W.}\ \bibnamefont {Chu}}, \bibinfo
  {author} {\bibfnamefont {R.}~\bibnamefont {Sankar}}, \bibinfo {author}
  {\bibfnamefont {F.~C.}\ \bibnamefont {Chou}}, \bibinfo {author}
  {\bibfnamefont {H.~T.}\ \bibnamefont {Jeng}}, \bibinfo {author}
  {\bibfnamefont {C.~S.}\ \bibnamefont {Chang}}, \ and\ \bibinfo {author}
  {\bibfnamefont {T.~M.}\ \bibnamefont {Chuang}},\ }\href@noop {} {\bibfield
  {journal} {\bibinfo  {journal} {Science Advances}\ }\textbf {\bibinfo
  {volume} {2}},\ \bibinfo {pages} {8} (\bibinfo {year} {2016})}\BibitemShut
  {NoStop}%
\bibitem [{\citenamefont {Matthias}\ \emph {et~al.}(1963)\citenamefont
  {Matthias}, \citenamefont {Geballe},\ and\ \citenamefont
  {Compton}}]{Matthia63}%
  \BibitemOpen
  \bibfield  {author} {\bibinfo {author} {\bibfnamefont {B.~T.}\ \bibnamefont
  {Matthias}}, \bibinfo {author} {\bibfnamefont {T.~H.}\ \bibnamefont
  {Geballe}}, \ and\ \bibinfo {author} {\bibfnamefont {V.~B.}\ \bibnamefont
  {Compton}},\ }\href@noop {} {\bibfield  {journal} {\bibinfo  {journal} {Rev.
  Mod. Phys.}\ }\textbf {\bibinfo {volume} {35}},\ \bibinfo {pages} {1}
  (\bibinfo {year} {1963})}\BibitemShut {NoStop}%
\bibitem [{\citenamefont {Okamoto}(1994)}]{Okamoto94}%
  \BibitemOpen
  \bibfield  {author} {\bibinfo {author} {\bibfnamefont {H.}~\bibnamefont
  {Okamoto}},\ }\href@noop {} {\bibfield  {journal} {\bibinfo  {journal} {J.
  Phase Equilib.}\ }\textbf {\bibinfo {volume} {15}},\ \bibinfo {pages} {191}
  (\bibinfo {year} {1994})}\BibitemShut {NoStop}%
\bibitem [{\citenamefont {Xu}\ \emph {et~al.}(1992)\citenamefont {Xu},
  \citenamefont {de~Groot},\ and\ \citenamefont {van~der Lugt}}]{Xu92}%
  \BibitemOpen
  \bibfield  {author} {\bibinfo {author} {\bibfnamefont {R.}~\bibnamefont
  {Xu}}, \bibinfo {author} {\bibfnamefont {A.~R.}\ \bibnamefont {de~Groot}}, \
  and\ \bibinfo {author} {\bibfnamefont {W.}~\bibnamefont {van~der Lugt}},\
  }\href@noop {} {\bibfield  {journal} {\bibinfo  {journal} {J. Phys. Condens.
  Matter}\ }\textbf {\bibinfo {volume} {4}},\ \bibinfo {pages} {2389} (\bibinfo
  {year} {1992})}\BibitemShut {NoStop}%
\bibitem [{\citenamefont {Shein}\ and\ \citenamefont
  {Ivanovskii}(2013)}]{Shein13}%
  \BibitemOpen
  \bibfield  {author} {\bibinfo {author} {\bibfnamefont {I.~R.}\ \bibnamefont
  {Shein}}\ and\ \bibinfo {author} {\bibfnamefont {A.~L.}\ \bibnamefont
  {Ivanovskii}},\ }\href@noop {} {\bibfield  {journal} {\bibinfo  {journal} {J.
  Supercond. Nov. Magn.}\ }\textbf {\bibinfo {volume} {26}},\ \bibinfo {pages}
  {1} (\bibinfo {year} {2013})}\BibitemShut {NoStop}%
\bibitem [{\citenamefont {Bonalde}\ \emph {et~al.}(2000)\citenamefont
  {Bonalde}, \citenamefont {Yanoff}, \citenamefont {Salamon}, \citenamefont
  {Van~Harlingen}, \citenamefont {Chia}, \citenamefont {Mao},\ and\
  \citenamefont {Maeno}}]{Bonalde2000}%
  \BibitemOpen
  \bibfield  {author} {\bibinfo {author} {\bibfnamefont {I.}~\bibnamefont
  {Bonalde}}, \bibinfo {author} {\bibfnamefont {B.~D.}\ \bibnamefont {Yanoff}},
  \bibinfo {author} {\bibfnamefont {M.~B.}\ \bibnamefont {Salamon}}, \bibinfo
  {author} {\bibfnamefont {D.~J.}\ \bibnamefont {Van~Harlingen}}, \bibinfo
  {author} {\bibfnamefont {E.~M.~E.}\ \bibnamefont {Chia}}, \bibinfo {author}
  {\bibfnamefont {Z.~Q.}\ \bibnamefont {Mao}}, \ and\ \bibinfo {author}
  {\bibfnamefont {Y.}~\bibnamefont {Maeno}},\ }\href@noop {} {\bibfield
  {journal} {\bibinfo  {journal} {Phys. Rev. Lett.}\ }\textbf {\bibinfo
  {volume} {85}},\ \bibinfo {pages} {4775} (\bibinfo {year}
  {2000})}\BibitemShut {NoStop}%
\bibitem [{\citenamefont {Chia}\ \emph
  {et~al.}(2003{\natexlab{a}})\citenamefont {Chia}, \citenamefont {Salamon},
  \citenamefont {Sugawara},\ and\ \citenamefont {Sato}}]{Chia03PRL}%
  \BibitemOpen
  \bibfield  {author} {\bibinfo {author} {\bibfnamefont {E.~E.~M.}\
  \bibnamefont {Chia}}, \bibinfo {author} {\bibfnamefont {M.~B.}\ \bibnamefont
  {Salamon}}, \bibinfo {author} {\bibfnamefont {H.}~\bibnamefont {Sugawara}}, \
  and\ \bibinfo {author} {\bibfnamefont {H.}~\bibnamefont {Sato}},\ }\href@noop
  {} {\bibfield  {journal} {\bibinfo  {journal} {Phys. Rev. Lett.}\ }\textbf
  {\bibinfo {volume} {91}},\ \bibinfo {pages} {247003} (\bibinfo {year}
  {2003}{\natexlab{a}})}\BibitemShut {NoStop}%
\bibitem [{\citenamefont {Chia}\ \emph {et~al.}(2004)\citenamefont {Chia},
  \citenamefont {Salamon}, \citenamefont {Sugawara},\ and\ \citenamefont
  {Sato}}]{Chia04}%
  \BibitemOpen
  \bibfield  {author} {\bibinfo {author} {\bibfnamefont {E.~E.~M.}\
  \bibnamefont {Chia}}, \bibinfo {author} {\bibfnamefont {M.~B.}\ \bibnamefont
  {Salamon}}, \bibinfo {author} {\bibfnamefont {H.}~\bibnamefont {Sugawara}}, \
  and\ \bibinfo {author} {\bibfnamefont {H.}~\bibnamefont {Sato}},\ }\href@noop
  {} {\bibfield  {journal} {\bibinfo  {journal} {Phys. Rev. B}\ }\textbf
  {\bibinfo {volume} {69}},\ \bibinfo {pages} {180509(R)} (\bibinfo {year}
  {2004})}\BibitemShut {NoStop}%
\bibitem [{\citenamefont {Chia}\ \emph {et~al.}(2005)\citenamefont {Chia},
  \citenamefont {Vandervelde}, \citenamefont {Salamon}, \citenamefont
  {Kikuchi}, \citenamefont {Sugawara},\ and\ \citenamefont {Sato}}]{Chia05}%
  \BibitemOpen
  \bibfield  {author} {\bibinfo {author} {\bibfnamefont {E.~E.~M.}\
  \bibnamefont {Chia}}, \bibinfo {author} {\bibfnamefont {D.}~\bibnamefont
  {Vandervelde}}, \bibinfo {author} {\bibfnamefont {M.~B.}\ \bibnamefont
  {Salamon}}, \bibinfo {author} {\bibfnamefont {D.}~\bibnamefont {Kikuchi}},
  \bibinfo {author} {\bibfnamefont {H.}~\bibnamefont {Sugawara}}, \ and\
  \bibinfo {author} {\bibfnamefont {H.}~\bibnamefont {Sato}},\ }\href@noop {}
  {\bibfield  {journal} {\bibinfo  {journal} {J. Phys.: Condens. Matter}\
  }\textbf {\bibinfo {volume} {17}},\ \bibinfo {pages} {L303} (\bibinfo {year}
  {2005})}\BibitemShut {NoStop}%
\bibitem [{\citenamefont {Hashimoto}\ \emph {et~al.}(2012)\citenamefont
  {Hashimoto}, \citenamefont {Cho}, \citenamefont {Shibauchi}, \citenamefont
  {Kasahara}, \citenamefont {Mizukami}, \citenamefont {Katsumata},
  \citenamefont {Tsuruhara}, \citenamefont {Terashima}, \citenamefont {Ikeda},
  \citenamefont {Tanatar}, \citenamefont {Kitano}, \citenamefont {Salovich},
  \citenamefont {Giannetta}, \citenamefont {Walmsley}, \citenamefont
  {Carrington}, \citenamefont {Prozorov},\ and\ \citenamefont
  {Matsuda}}]{Matsuda12}%
  \BibitemOpen
  \bibfield  {author} {\bibinfo {author} {\bibfnamefont {K.}~\bibnamefont
  {Hashimoto}}, \bibinfo {author} {\bibfnamefont {K.}~\bibnamefont {Cho}},
  \bibinfo {author} {\bibfnamefont {T.}~\bibnamefont {Shibauchi}}, \bibinfo
  {author} {\bibfnamefont {S.}~\bibnamefont {Kasahara}}, \bibinfo {author}
  {\bibfnamefont {Y.}~\bibnamefont {Mizukami}}, \bibinfo {author}
  {\bibfnamefont {R.}~\bibnamefont {Katsumata}}, \bibinfo {author}
  {\bibfnamefont {Y.}~\bibnamefont {Tsuruhara}}, \bibinfo {author}
  {\bibfnamefont {T.}~\bibnamefont {Terashima}}, \bibinfo {author}
  {\bibfnamefont {H.}~\bibnamefont {Ikeda}}, \bibinfo {author} {\bibfnamefont
  {M.~A.}\ \bibnamefont {Tanatar}}, \bibinfo {author} {\bibfnamefont
  {H.}~\bibnamefont {Kitano}}, \bibinfo {author} {\bibfnamefont
  {N.}~\bibnamefont {Salovich}}, \bibinfo {author} {\bibfnamefont {R.~W.}\
  \bibnamefont {Giannetta}}, \bibinfo {author} {\bibfnamefont {P.}~\bibnamefont
  {Walmsley}}, \bibinfo {author} {\bibfnamefont {A.}~\bibnamefont
  {Carrington}}, \bibinfo {author} {\bibfnamefont {R.}~\bibnamefont
  {Prozorov}}, \ and\ \bibinfo {author} {\bibfnamefont {Y.}~\bibnamefont
  {Matsuda}},\ }\href@noop {} {\bibfield  {journal} {\bibinfo  {journal}
  {Science}\ }\textbf {\bibinfo {volume} {336}},\ \bibinfo {pages} {1554}
  (\bibinfo {year} {2012})}\BibitemShut {NoStop}%
\bibitem [{\citenamefont {Ortenzi}\ \emph {et~al.}(2015)\citenamefont
  {Ortenzi}, \citenamefont {Gretarsson}, \citenamefont {Kasahara},
  \citenamefont {Matsuda}, \citenamefont {Shibauchi}, \citenamefont
  {Finkelstein}, \citenamefont {Wu}, \citenamefont {Julian}, \citenamefont
  {Kim}, \citenamefont {Mazin},\ and\ \citenamefont {Boeri}}]{Matsuda15}%
  \BibitemOpen
  \bibfield  {author} {\bibinfo {author} {\bibfnamefont {L.}~\bibnamefont
  {Ortenzi}}, \bibinfo {author} {\bibfnamefont {H.}~\bibnamefont {Gretarsson}},
  \bibinfo {author} {\bibfnamefont {S.}~\bibnamefont {Kasahara}}, \bibinfo
  {author} {\bibfnamefont {Y.}~\bibnamefont {Matsuda}}, \bibinfo {author}
  {\bibfnamefont {T.}~\bibnamefont {Shibauchi}}, \bibinfo {author}
  {\bibfnamefont {K.~D.}\ \bibnamefont {Finkelstein}}, \bibinfo {author}
  {\bibfnamefont {W.}~\bibnamefont {Wu}}, \bibinfo {author} {\bibfnamefont
  {S.~R.}\ \bibnamefont {Julian}}, \bibinfo {author} {\bibfnamefont {Y.-J.}\
  \bibnamefont {Kim}}, \bibinfo {author} {\bibfnamefont {I.~I.}\ \bibnamefont
  {Mazin}}, \ and\ \bibinfo {author} {\bibfnamefont {L.}~\bibnamefont
  {Boeri}},\ }\href@noop {} {\bibfield  {journal} {\bibinfo  {journal} {Phys.
  Rev. Lett.}\ }\textbf {\bibinfo {volume} {114}},\ \bibinfo {pages} {047001}
  (\bibinfo {year} {2015})}\BibitemShut {NoStop}%
\bibitem [{\citenamefont {Cho}\ \emph {et~al.}(2012)\citenamefont {Cho},
  \citenamefont {Tanatar}, \citenamefont {Spyrison}, \citenamefont {Kim},
  \citenamefont {Song}, \citenamefont {Dai}, \citenamefont {Zhang},\ and\
  \citenamefont {Prozorov}}]{Prozorov12}%
  \BibitemOpen
  \bibfield  {author} {\bibinfo {author} {\bibfnamefont {K.}~\bibnamefont
  {Cho}}, \bibinfo {author} {\bibfnamefont {M.~A.}\ \bibnamefont {Tanatar}},
  \bibinfo {author} {\bibfnamefont {N.}~\bibnamefont {Spyrison}}, \bibinfo
  {author} {\bibfnamefont {H.}~\bibnamefont {Kim}}, \bibinfo {author}
  {\bibfnamefont {Y.}~\bibnamefont {Song}}, \bibinfo {author} {\bibfnamefont
  {P.~C.}\ \bibnamefont {Dai}}, \bibinfo {author} {\bibfnamefont {C.~L.}\
  \bibnamefont {Zhang}}, \ and\ \bibinfo {author} {\bibfnamefont
  {R.}~\bibnamefont {Prozorov}},\ }\href@noop {} {\bibfield  {journal}
  {\bibinfo  {journal} {Phys. Rev. B}\ }\textbf {\bibinfo {volume} {86}},\
  \bibinfo {pages} {5} (\bibinfo {year} {2012})}\BibitemShut {NoStop}%
\bibitem [{\citenamefont {Prozorov}\ \emph {et~al.}(2000)\citenamefont
  {Prozorov}, \citenamefont {Giannetta}, \citenamefont {Carrington},\ and\
  \citenamefont {Araujo-Moreira}}]{Prozorov00}%
  \BibitemOpen
  \bibfield  {author} {\bibinfo {author} {\bibfnamefont {R.}~\bibnamefont
  {Prozorov}}, \bibinfo {author} {\bibfnamefont {R.~W.}\ \bibnamefont
  {Giannetta}}, \bibinfo {author} {\bibfnamefont {A.}~\bibnamefont
  {Carrington}}, \ and\ \bibinfo {author} {\bibfnamefont {F.~M.}\ \bibnamefont
  {Araujo-Moreira}},\ }\href@noop {} {\bibfield  {journal} {\bibinfo  {journal}
  {Phys. Rev. B}\ }\textbf {\bibinfo {volume} {62}},\ \bibinfo {pages} {115}
  (\bibinfo {year} {2000})}\BibitemShut {NoStop}%
\bibitem [{\citenamefont {Prozorov}\ \emph {et~al.}(2001)\citenamefont
  {Prozorov}, \citenamefont {Giannetta}, \citenamefont {Schlueter},
  \citenamefont {Kini}, \citenamefont {Mohtasham}, \citenamefont {Winter},\
  and\ \citenamefont {Gard}}]{Prozorov01}%
  \BibitemOpen
  \bibfield  {author} {\bibinfo {author} {\bibfnamefont {R.}~\bibnamefont
  {Prozorov}}, \bibinfo {author} {\bibfnamefont {R.~W.}\ \bibnamefont
  {Giannetta}}, \bibinfo {author} {\bibfnamefont {J.}~\bibnamefont
  {Schlueter}}, \bibinfo {author} {\bibfnamefont {A.~M.}\ \bibnamefont {Kini}},
  \bibinfo {author} {\bibfnamefont {J.}~\bibnamefont {Mohtasham}}, \bibinfo
  {author} {\bibfnamefont {R.~W.}\ \bibnamefont {Winter}}, \ and\ \bibinfo
  {author} {\bibfnamefont {G.~L.}\ \bibnamefont {Gard}},\ }\href@noop {}
  {\bibfield  {journal} {\bibinfo  {journal} {Phys. Rev. B}\ }\textbf {\bibinfo
  {volume} {63}},\ \bibinfo {pages} {052506} (\bibinfo {year}
  {2001})}\BibitemShut {NoStop}%
\bibitem [{\citenamefont {Chia}\ \emph
  {et~al.}(2003{\natexlab{b}})\citenamefont {Chia}, \citenamefont
  {Van~Harlingen}, \citenamefont {Salamon}, \citenamefont {Yanoff},
  \citenamefont {Bonalde},\ and\ \citenamefont {Sarrao}}]{Chia03}%
  \BibitemOpen
  \bibfield  {author} {\bibinfo {author} {\bibfnamefont {E.~E.~M.}\
  \bibnamefont {Chia}}, \bibinfo {author} {\bibfnamefont {D.~J.}\ \bibnamefont
  {Van~Harlingen}}, \bibinfo {author} {\bibfnamefont {M.~B.}\ \bibnamefont
  {Salamon}}, \bibinfo {author} {\bibfnamefont {B.~D.}\ \bibnamefont {Yanoff}},
  \bibinfo {author} {\bibfnamefont {I.}~\bibnamefont {Bonalde}}, \ and\
  \bibinfo {author} {\bibfnamefont {J.~L.}\ \bibnamefont {Sarrao}},\
  }\href@noop {} {\bibfield  {journal} {\bibinfo  {journal} {Phys. Rev. B}\
  }\textbf {\bibinfo {volume} {67}},\ \bibinfo {pages} {014527} (\bibinfo
  {year} {2003}{\natexlab{b}})}\BibitemShut {NoStop}%
\bibitem [{\citenamefont {Carrington}\ \emph {et~al.}(1999)\citenamefont
  {Carrington}, \citenamefont {Bonalde}, \citenamefont {Prozorov},
  \citenamefont {Giannetta}, \citenamefont {Kini}, \citenamefont {Schlueter},
  \citenamefont {Wang}, \citenamefont {Geiser},\ and\ \citenamefont
  {Williams}}]{Carrington1999}%
  \BibitemOpen
  \bibfield  {author} {\bibinfo {author} {\bibfnamefont {A.}~\bibnamefont
  {Carrington}}, \bibinfo {author} {\bibfnamefont {I.~J.}\ \bibnamefont
  {Bonalde}}, \bibinfo {author} {\bibfnamefont {R.}~\bibnamefont {Prozorov}},
  \bibinfo {author} {\bibfnamefont {R.~W.}\ \bibnamefont {Giannetta}}, \bibinfo
  {author} {\bibfnamefont {A.~M.}\ \bibnamefont {Kini}}, \bibinfo {author}
  {\bibfnamefont {J.}~\bibnamefont {Schlueter}}, \bibinfo {author}
  {\bibfnamefont {H.~H.}\ \bibnamefont {Wang}}, \bibinfo {author}
  {\bibfnamefont {U.}~\bibnamefont {Geiser}}, \ and\ \bibinfo {author}
  {\bibfnamefont {J.~M.}\ \bibnamefont {Williams}},\ }\href@noop {} {\bibfield
  {journal} {\bibinfo  {journal} {Phys. Rev. Lett.}\ }\textbf {\bibinfo
  {volume} {83}},\ \bibinfo {pages} {4172} (\bibinfo {year}
  {1999})}\BibitemShut {NoStop}%
\bibitem [{\citenamefont {Muhlschlegel}(1959)}]{derivation59}%
  \BibitemOpen
  \bibfield  {author} {\bibinfo {author} {\bibfnamefont {B.}~\bibnamefont
  {Muhlschlegel}},\ }\href@noop {} {\bibfield  {journal} {\bibinfo  {journal}
  {Zeitschrift F$\ddot{u}$r Physik}\ }\textbf {\bibinfo {volume} {155}},\
  \bibinfo {pages} {313} (\bibinfo {year} {1959})}\BibitemShut {NoStop}%
\bibitem [{\citenamefont {Prozorov}\ \emph {et~al.}(2006)\citenamefont
  {Prozorov}, \citenamefont {Olheiser}, \citenamefont {Giannetta},
  \citenamefont {Uozato},\ and\ \citenamefont {Tamegai}}]{ProzorovlowT}%
  \BibitemOpen
  \bibfield  {author} {\bibinfo {author} {\bibfnamefont {R.}~\bibnamefont
  {Prozorov}}, \bibinfo {author} {\bibfnamefont {T.~A.}\ \bibnamefont
  {Olheiser}}, \bibinfo {author} {\bibfnamefont {R.~W.}\ \bibnamefont
  {Giannetta}}, \bibinfo {author} {\bibfnamefont {K.}~\bibnamefont {Uozato}}, \
  and\ \bibinfo {author} {\bibfnamefont {T.}~\bibnamefont {Tamegai}},\
  }\href@noop {} {\bibfield  {journal} {\bibinfo  {journal} {Phys. Rev. B}\
  }\textbf {\bibinfo {volume} {73}},\ \bibinfo {pages} {184523} (\bibinfo
  {year} {2006})}\BibitemShut {NoStop}%
\bibitem [{\citenamefont {Tinkham}(1975)}]{Tinkham}%
  \BibitemOpen
  \bibfield  {author} {\bibinfo {author} {\bibfnamefont {M.}~\bibnamefont
  {Tinkham}},\ }\href@noop {} {\bibfield  {journal} {\bibinfo  {journal}
  {Introduction to Superconductivity, McGraw-Hill, New York}\ } (\bibinfo
  {year} {1975})}\BibitemShut {NoStop}%
\bibitem [{\citenamefont {Gross}\ \emph {et~al.}(1986)\citenamefont {Gross},
  \citenamefont {Chandrasekhar}, \citenamefont {Einzel}, \citenamefont
  {Andres}, \citenamefont {Hirschfeld}, \citenamefont {Ott}, \citenamefont
  {Beuers}, \citenamefont {Fisk},\ and\ \citenamefont {Smith}}]{Gross86}%
  \BibitemOpen
  \bibfield  {author} {\bibinfo {author} {\bibfnamefont {F.}~\bibnamefont
  {Gross}}, \bibinfo {author} {\bibfnamefont {B.~S.}\ \bibnamefont
  {Chandrasekhar}}, \bibinfo {author} {\bibfnamefont {D.}~\bibnamefont
  {Einzel}}, \bibinfo {author} {\bibfnamefont {K.}~\bibnamefont {Andres}},
  \bibinfo {author} {\bibfnamefont {P.~J.}\ \bibnamefont {Hirschfeld}},
  \bibinfo {author} {\bibfnamefont {H.~R.}\ \bibnamefont {Ott}}, \bibinfo
  {author} {\bibfnamefont {J.}~\bibnamefont {Beuers}}, \bibinfo {author}
  {\bibfnamefont {Z.}~\bibnamefont {Fisk}}, \ and\ \bibinfo {author}
  {\bibfnamefont {J.~L.}\ \bibnamefont {Smith}},\ }\href@noop {} {\bibfield
  {journal} {\bibinfo  {journal} {Zeitschrift f{\"u}r Physik B Condensed
  Matter}\ }\textbf {\bibinfo {volume} {64}},\ \bibinfo {pages} {175} (\bibinfo
  {year} {1986})}\BibitemShut {NoStop}%
\bibitem [{\citenamefont {Kresin}\ and\ \citenamefont
  {Parkhomenko}(1975)}]{Kresin75}%
  \BibitemOpen
  \bibfield  {author} {\bibinfo {author} {\bibfnamefont {V.}~\bibnamefont
  {Kresin}}\ and\ \bibinfo {author} {\bibfnamefont {V.}~\bibnamefont
  {Parkhomenko}},\ }\href@noop {} {\bibfield  {journal} {\bibinfo  {journal}
  {Sov. Phys. Solid State}\ }\textbf {\bibinfo {volume} {16}},\ \bibinfo
  {pages} {2180} (\bibinfo {year} {1975})}\BibitemShut {NoStop}%
\bibitem [{\citenamefont {Orlando}\ \emph {et~al.}(1979)\citenamefont
  {Orlando}, \citenamefont {McNiff}, \citenamefont {Foner},\ and\ \citenamefont
  {Beasley}}]{Orlando79}%
  \BibitemOpen
  \bibfield  {author} {\bibinfo {author} {\bibfnamefont {T.~P.}\ \bibnamefont
  {Orlando}}, \bibinfo {author} {\bibfnamefont {E.~J.}\ \bibnamefont {McNiff}},
  \bibinfo {author} {\bibfnamefont {S.}~\bibnamefont {Foner}}, \ and\ \bibinfo
  {author} {\bibfnamefont {M.~R.}\ \bibnamefont {Beasley}},\ }\href@noop {}
  {\bibfield  {journal} {\bibinfo  {journal} {Phys. Rev. B}\ }\textbf {\bibinfo
  {volume} {19}},\ \bibinfo {pages} {4545} (\bibinfo {year}
  {1979})}\BibitemShut {NoStop}%
\bibitem [{\citenamefont {et~al.}(shed)}]{Zhu17}%
  \BibitemOpen
  \bibfield  {author} {\bibinfo {author} {\bibfnamefont {J.-X.~Z.}\
  \bibnamefont {et~al.}},\ }\href@noop {} {\  (\bibinfo {year} {to be
  published})}\BibitemShut {NoStop}%
\end{thebibliography}%
\bigskip

\end{document}